# Atomtronics

Luigi Amico[1,2] and Malcolm G. Boshier[3]

[1] CNR-MATIS-IMM, Dipartimento di Fisica Universita' di Catania & INFN Laboratori Nazionali del Sud, via S. Sofia 64, 95127 Catania (Italy).
[2] Center for Quantum Technologies, National University of Singapore, 3 Science Drive 2, 117543 Singapore.
[3] Physics Division, Los Alamos National Laboratory, Los Alamos, NM 87545, USA.

**Status** – Atomtronics is an emerging field seeking to realize atomic circuits exploiting ultra-cold atoms manipulated in micro-magnetic or laser-generated micro-optical or circuits [1,2]. Compared with the conventional electronics, the concept of Atomtronics has several key aspects. First, current-flows in Atomtronics circuits are made of neutrally charged carriers put in motion by applying a 'potential drop' induced by a bias in the chemical potential or by phase imprinting techniques, or by stirring the atomic gas with a laser beam 'tea spoon'. Second, the carriers in the circuit can be of different physical nature, i.e. bosons, fermions, or a mixture of thereof, and with mutual interaction that can be tuned from short to long-range, from attractive to repulsive etc. Finally, quantum coherence is a characteristic trait of the systems harnessed in the circuit.

The typically low decoherence/dissipation rates of cold atoms systems and the high controllability and enhanced flexibility of potentials for ultracold matter make Atomtronics very attractive for enlarging the scope of existing cold-atom quantum technology [3]. Both technological and fundamental issues in physics can be addressed.

Elementary Atomtronic circuits have already been realized, mimicking both conventional electronics like diodes, PNP junctions [1,4] and elements of quantum electronics such as Josephson junctions SQUID's [5,6,7]. Atomtronics, however, is not strictly limited to developing electronic-like components: it aims at providing new concepts of quantum devices, integrated in a circuitry that may be of a radically new type. A remarkable impact in diverse subfields of quantum technology, including quantum metrology, quantum information, and quantum computation, is expected. Indeed, interferometric high precision sensors with matter waves promise a considerable gain in sensitivity compared with the existing solutions (for rotation sensing, in particular, the gaining factor to the light based technology, can be up to $\sim 10^{10}$, for equal enclosed areas) [8]. New hybrid cold-atom/solid state systems have been conceived, both to assemble devices with enhanced quantum coherence and to develop a new avenue for the diagnostics of the interfaced systems [9,10].

With atomic flows and high flexibility in the confinement geometry offered by Atomtronics, several problems that cannot even be defined within standard quantum simulator architectures could be addressed [3]. The situation is comparable to the development of solid state physics: a fruitful avenue in that field is to study the (electronic) current in the system as a response to an external perturbation. With the same logic, key issues in many-body physics, like frustration effects, topological constraints, quantum Hall edge currents etc., can be addressed by Atomtronics, with unprecedented flexibility in parameters.

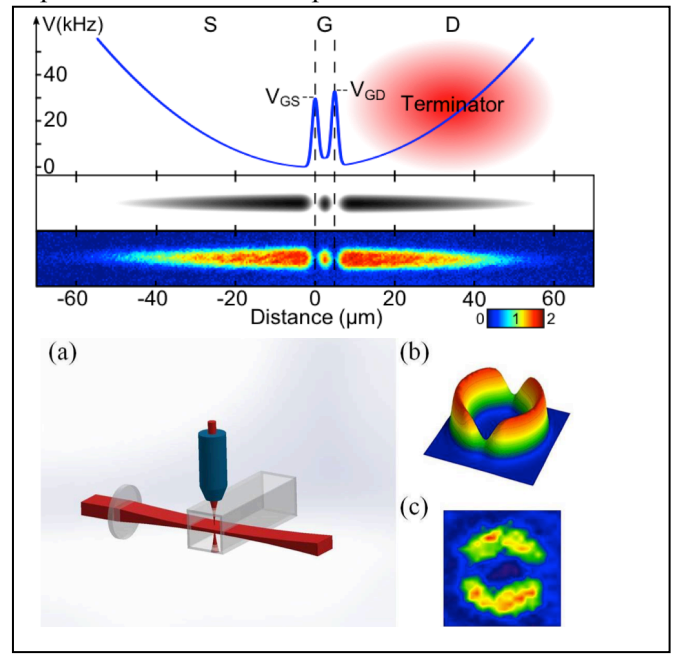

Figure 1 – Upper panel: The atomtronic transistor. Source, Gate, and Drain wells labeled S, G, and D, respectively are created with an hybrid magnetic and optical potential. The middle panel is a calculated potential energy density plot while the lower panel shows a false color in-trap absorption image of atoms occupying all three wells. An optical density scale is shown below the horizontal axis [From [2]].
Lower panels: The dc-Atomtronics Quantum Inteference Device (AQUID) (b), (c) realized with painted potential techniques (a) [From[7]].

**Current and Future Challenges** – Although several proof-of-principle schemes to assemble Atomtronics integrated circuits have been carried out, specific protocols to design and benchmark elementary circuits are still under discussion. Flexible-geometry wave guides, matter wave beam splitters, 'inductors' etc. need to be tailored, and specific schemes for assembling the conceived circuital elements need to be provided (see [11,12,13] for recent design approaches). It would be desirable to work out a heuristic approach to the circuitry, leading, for example, to lumped parameters models analogous to conventional electronics.

In metrology, it is certainly interesting to realize structures with extended storage times with concomitant improvement in measurement precision (see Fig.2).

Atomtronics has the potential to provide a new platform for quantum signal processing and a new quantum classical interface. Combining some of the virtues of Josephson junctions flux qubits (such as macroscopic quantum coherence) with cold atom advantages (reduced decoherence), 'atomic flux qubits' can open up a new direction in physical implementations of quantum computation. A specific avenue to such goals is to analyze a system of quantum degenerate particles confined in ring shaped potentials with few localized weak links. Although the emergence of a qubit dynamics of clockwise and anticlockwise atomic coherent flows in a mesoscopic ring lattice of BECs has been predicted in realistic situations[14], Rabi oscillations have not yet been observed experimentally. Also, protocols have been provided for tunable ring-ring (qubit-qubit) coupling quantum devices. Quantum gates remain to be devised.

Atomtronics is instrumental for devising new concepts for quantum simulation. Several ramifications can be envisaged. Issues in quantum material science like topological matter, spin liquids etc. can be studied, analyzing the flowing current responding to 'potential drops'. Exploiting the power of ultracold gases for studying far from equilibrium systems [15], non-equilibrium statistical mechanics and quench dynamics could be addressed by inspecting the dynamics of atomic flows in different geometries.

It is intriguing to explore physical systems in mesoscopic regimes. Any configuration (compatible with the spatial resolution of the Atomtronic circuit) can be studied, in principle. Finally, interfacing cold-atom and solid state systems in hybrid structures is certainly one of the most interesting avenues to follow. From one side, cold atoms could be utilized to diagnose the proximal solid state/mesoscopic system. On the other hand, the hybrid platform could define quantum circuits with enhanced coherence time, where quantum states are are stored and exchanged between the solid state and the cold atom systems [9].

**Advances in Science and Technology to Meet Challenges –** The accelerating growth of Atomtronics is in part due to recent progress in optical microfabrication. That technology allows central issues of cold atom systems, like scalability, reconfigurability, and stability to be feasibly addressed [2]. In many current and envisaged investigations, there is a need to push for further miniaturization of the circuits. The current lower limit here is generically imposed by the diffraction limit of the employed optics. With the current technologies, the different confining potentials (in shape and intensities) can be controlled on the micrometer spatial ranges, achieving a *5% rms* smoothness. Going to the sub-micron regime, although challenging, might be accessible in the near future. A milestone, in this regime, would be to simultaneously optimize laser and vacuum technology as well as optical delivery systems. Even at the current spatial scale, mesoscopic quantum effects could be accessed. In particular, ring-shaped potentials of tens of micrometer radii have great potentials both for practical and fundamental research, for instance, as quantum memories and quantum simulators. The scalability of multiple-ring structures will be certainly fostered by tailoring optical potentials beyond the Laguerre-Gauss type (e.g. employing Bessel-Gauss laser beams). Challenges such as the detection of 'cat states' and the creation of ring-ring interactions are expected to be reduced by modulating the confinement along the ring (or ring lattice) potentials.

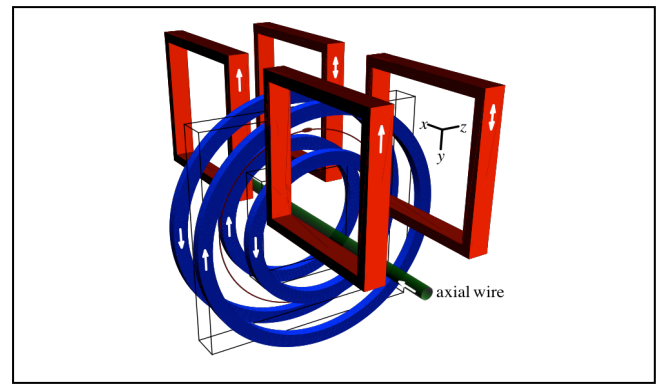

Figure 2 – The large storage vertical ring for Bose-Einstein condensates realized by the Arnold group (average diam eter 10 cm). [From[8]]

For most, if not all, applications, it would be useful to achieve better coherence times (>1s). For high precision interferometry, higher repetition rates (>$1Hz$), with larger thermal/condensed atom number (>$10^6/10^9$) with sufficiently low densities are important.

A central issue for creating complex Atomtronic integrated circuits is minimizing the operating time on the circuit and speeding up communication among different circuit elements. Currently, typical time scales are in the *ms* range, but a thorough analysis of the parameters and physics controlling time scales is still missing. A seemingly feasible perspective, provided by such new quantum technology, is to create schemes in which components and connectors can be changed dynamically in the course of the circuit life [12]. In this context, it would be important to study to what extent chemical potential and other effective 'potential drops' changes can be detected as a function of time.

Finally, novel avenues for diagnostics of the Atomtronic system's current state, beyond the current absortion imaging technique, should be developed. Non-destructive techniques will be particularly useful.

**Concluding Remarks –** Atomtronics is expected to provide novel designs for quantum devices exploiting quantum phenomena such as superposition and entanglement and extend the scope of the quantum simulation. Although initial inspiration came from existing devices in solid state electronics, Atomtronics has the potential to define a new class of questions and answers in basic science and technology, complementing standard electronics and integrated optics. Prototypes of instruments for sensing, quantum gates, quantum memories functioning with cold atom currents in ring-shaped architectures, and realizing data busses seem to be accessible near-term goals. An improved understanding of real electronic systems may also be achieved.

It appears very likely that Atomtronics will contribute to breakthrough technology developments in the years to come. To enhance knowledge transfer between basic science and technology and to develop basic science into devices and instruments, it is highly desirable that industrial partners take part in this activity: the melting-pot that historically has been the core arena for scientific progress would thus be realized in this exciting new field.

---

[1]The Atomtronics literature is already large, but editorial constraints limit the bibliography to being representative rather than exhaustive.